# Magnetoelectric coupling and decoupling in multiferroic hexagonal YbFeO$_3$ thin films

*Yu Yun,[1][‡] Xin Li,[1][‡] Arashdeep Singh Thind,[2] Yuewei Yin,[1] Hao Liu,[3] Qiang Li,[3] Wenbin Wang,[3] Alpha T. N'Diaye,[4] Corbyn Mellinger,[1] Xuanyuan Jiang,[1] Rohan Mishra,[2,5] Xiaoshan Xu[1,6]\**

[1]Department of Physics and Astronomy, University of Nebraska, Lincoln, Nebraska 68588, USA

[2]Institute of Materials Science & Engineering, Washington University in St. Louis, St. Louis MO, USA

[3]Institute for Nanoelectronic Devices and Quantum Computing, Fudan University, Shanghai 200433, China

[4]Advanced Light Source, Lawrence Berkeley National Laboratory, Berkeley, California 94720, USA

[5]Department of Mechanical Engineering & Materials Science, Washington University in St. Louis, St. Louis MO, USA

[6]Nebraska Center for Materials and Nanoscience, University of Nebraska, Lincoln, Nebraska 68588, USA

[‡]These authors contributed equally to this work.

\*Corresponding Author: Email: xiaoshan.xu@unl.edu (X.X.)

## ABSTRACT

The coupling between ferroelectric and magnetic orders in multiferroic materials and the nature of magnetoelectric (ME) effects are enduring experimental challenges. In this work, we have studied the response of magnetization to ferroelectric switching in thin-film hexagonal YbFeO$_3$, a prototypical improper multiferroic. The bulk ME decoupling and potential domain-wall ME coupling were revealed using x-ray magnetic circular dichroism (XMCD) measurements with in-situ ferroelectric polarization switching. Our Landau theory analysis suggests that the bulk ME-coupled ferroelectric switching path has a higher energy barrier than that of the ME-decoupled path; this extra barrier energy is also too high to be reduced by the magneto-static energy in the process of breaking single magnetic domains into multi-domains. In addition, the reduction of magnetization around the ferroelectric domain walls predicted by the Landau theory may induce the domain-wall ME coupling in which the magnetization is correlated with the density of ferroelectric domain walls. These results provide important experimental evidence and theoretical insights into the rich possibilities of ME couplings in hexagonal ferrites, such as manipulating the magnetic states by an electric field.

**INTRODUCTION**

The coupling of ferroelectric (FE) and magnetic orders in multiferroic materials is attractive due to its fundamental importance as well as promising applications in electronic devices [1,2,3,4]. The holy grail in a single-phase material is to achieve magnetoelectric (ME) coupling wherein the magnetization can be switched by applying an electric field above room temperature. Static ME coupling purely from electromagnetic interactions is unlikely since ferroelectric and magnetic orders follow different symmetric transformations [5,6,7]. On the other hand, the interplay between structural distortions, ferroelectricity, anti-ferromagnetism (AFM), and weak ferromagnetism (WFM) in complex materials like hexagonal ferrites (h-$R$FeO$_3$, $R$ = Y, Sc, Ho-Lu) allows for a rich set of coupling scenarios [8,9], which makes h-$R$FeO$_3$ an intriguing material family to investigate.

Under the ferroelectric state, hexagonal ferrites belong to non-centrosymmetric $P6_3cm$ space group at room temperature, similar to the isomorphic hexagonal $R$MnO$_3$ (h-$R$MnO$_3$). In this phase, a layer of FeO$_5$ bipyramids is separated by $R$ ions, and there exists an in-plane 60º rotation between the triangle of FeO$_5$ bipyramids in odd and even layers, as shown in Fig. 1a. The ferroelectricity of h-$R$FeO$_3$ is improper since the polarization is induced by the primary K$_3$ structural distortion, which is nonpolar. In the K$_3$ structural distortion, FeO$_5$ bipyramids tilt collectively, which is accompanied by the buckling of the $R$ layers. The induced vertical displacement of the atoms, corresponding to the $\Gamma_2^-$ structural distortion, directly generates the FE polarization of h-$R$FeO$_3$ [10, 11], as depicted in Fig. 1a for a negative polarization along the $c$ axis. The mostly in-plane Fe spins cant out-of-plane due to the single-ion magnetic anisotropy and the Dzyaloshinskii-Moriya (DM) interaction in h-$R$FeO$_3$, leading to WFM at low temperatures [12, 13] with the spontaneous magnetization along the $c$ axis, as illustrated in Fig. 1b. The anti-ferromagnetic exchange interactions between the Fe sites keep an in-plane 120º alignment between the Fe spins, as shown in Fig. 1c. What distinguishes h-$R$FeO$_3$ from h-$R$MnO$_3$, is the higher antiferromagnetic ordering temperature originating from the stronger Fe-Fe magnetic interactions and the WFM, which can be greatly amplified by the magnetic rare-earth centers such as Yb [12, 14].

The interplay among the structural distortion, FE order, AFM order, and WFM order, supports multiple possible scenarios in terms of the response of magnetic orders to the FE polarization switching in h-$R$FeO$_3$. For bulk ME coupling, it was argued that starting from a single magnetic domain, the polarization switching may trigger the formation of multiple magnetic domains to reduce the magneto-static energy and, of course, the magnetization, suggesting strong ME coupling [8]. For domain-wall ME coupling, it was demonstrated in h-$R$MnO$_3$ that the mismatched length scale of the FE and magnetic domain walls could magnetize the domain walls in contrast to the zero magnetization in the bulk [9, 15]. On the other hand, decoupled WFM and FE domains have recently been observed in h-(Lu,Sc)FeO$_3$ single crystals, which was explained by different topological behaviors between the polarization and magnetization switching, different energy scales for the FE and magnetic orders, and the corresponding different length scales between the FE and magnetic domains [16]. Nevertheless, neither the role of magneto-static energy in bulk ME coupling nor the ME coupling at the domain walls, especially during the ferroelectric switching process in h-$R$FeO$_3$ — multiferroics with a larger magneto-static energy—, remain open questions.

In this work, we focus on h-YbFeO$_3$ thin films, in which a larger scale of magneto-static energy is expected due to the larger magnetization on the Fe$^{3+}$ sites (0.03 $\mu_B$/Fe), which is further amplified by the Yb$^{3+}$ sites (1.6 $\mu_B$/Yb) [12], in contrast to the smaller saturation magnetization in h-LuFeO$_3$ (0.02 $\mu_B$/Fe), h-ScFeO$_3$ (0.015 $\mu_B$/Fe), and h-(Lu,Sc)FeO$_3$ (0.01 $\mu_B$/Fe) [14, 16, 17]. Despite the observation of much smaller magnetic domain size (~ 1 μm) consistent with the larger magneto-static energy, the in-situ measurements of the magnetization during the polarization reversal show decoupled FE polarization and magnetization switching. Our analysis based on a Landau theory model reveals that the bulk ME coupled switching (polarization switching accompanied by magnetization reversal) needs to overcome a higher energy barrier, with which the magneto-static energy is not large enough to disrupt the switching path even for h-YbFeO$_3$. The theoretical calculation also predicts suppressed magnetization at the FE domain walls, suggesting a ME coupling due to clamped antiferromagnetic domain walls. These results set a step forward for the exploration of ME coupling and decoupling in single-phase multiferroic h-YbFeO$_3$ for ME devices.

**RESULTS**

**Crystal structure characterization**

Bottom electrodes are essential to investigate the ME coupling by reversing FE polarization through the application of an electric field. The absence of lattice-matched bottom electrodes for epitaxial growth of hexagonal ferrites thin films has been a long-standing issue that limited the study of its ME coupling. Here, high-quality h-YbFeO$_3$/CFO/LSMO heterestructures were designed and fabricated on STO (111) substrates by PLD system, in which the LSMO (La$_{0.7}$Sr$_{0.3}$MnO$_3$) layer is used as the bottom electrode and the CFO (CoFe$_2$O$_4$) layer serves as a buffer layer to reduce the lattice mismatch between the LSMO layer and the h-YbFeO$_3$ layer, and stabilize the hexagonal structure of h-YbFeO$_3$. To illustrate the structure and phase of h-YbFeO$_3$ films, x-ray diffraction (XRD), reflection high energy electron diffraction (RHEED), and scanning transmission electron microscopy (STEM) measurements were carried out. Fig. 1d shows the representative $\theta$-$2\theta$ XRD scan of h-YbFeO$_3$/CFO/LSMO/STO heterostructure without obvious impurity structural phases. As shown in Fig. 1e, sharp and clear streaks indicate smooth surfaces and the epitaxial relationship between layers. The intense diffraction streaks separated by weak streaks in the RHEED pattern of h-YbFeO$_3$ are indicators of the $P6_3cm$ structure [10]. Fig. 1f shows a wide field-of-view high-angle annular dark-field (HAADF) image of the h-YbFeO$_3$/CFO/LSMO/STO heterostructure film. The K$_3$ structural distortion of h-YbFeO$_3$ can be identified through the buckling of the Yb layers, which gives rise to the ferroelectricity [18, 19]. As shown in the magnified HAADF image in Fig. 1f, FE polarization in this case is pointing up according to the periodical "two-up-one-down" distribution of the Yb atoms.

**MFM experiments**

As shown in Fig. 2a, the magnetic domains of h-YbFeO$_3$ films were investigated using magnetic force microscopy (MFM) at 12 K under an out-of-plane magnetic field varying from 10 kOe to -10 kOe (along blue arrow) and -10 kOe to 10 kOe (along red arrow). MFM measures the magnetic force on the tip, which is proportional to the gradient of the magnetic field and the magnetization of the tip (600 Oe coercive field). The gradient of the magnetic field could come from the inhomogeneous magnetization, i.e. magnetic domain structures on a flat surface, and from an uneven surface of a magnetic material. Comparing the MFM images in Fig. 2a and the surface topography in Fig. 2b, for the top and bottom edges of the image, the MFM contrast most likely comes from the island-like surface topography (defect sites, labeled by white circles in Fig. 2a and Fig. 2b). On the other hand, the contrast of flat center parts (dashed squares) in MFM images reflects the magnetic domain structures (red circle in Fig. 2a) of the h-YbFeO$_3$ films, which will be discussed below.

Starting at 10 kOe in Fig. 2a (along blue arrow), the MFM image shows minimal contrast, illustrating a mostly homogeneous distribution of the saturated magnetization associated with a magnetic single-domain state. From 10 kOe to -2 kOe, no obvious change in MFM images was observed, suggesting that the magnetization was not switched, which is consistent with previous observations [12, 20]. As the magnetic field was changed to -2 kOe, a larger contrast was identified in the center of the MFM image, illustrating the appearance of multiple magnetic domains or partial reversal of magnetization along the *c* axis driven by the Zeeman energy. As the magnetic field further decreased to -4 kOe, the contrast of the MFM image reached a maximum, and as the magnetic field reached -6 kOe, the contrast of the MFM image decreased compared with that in the image taken at -4 kOe, demonstrating further reversal of magnetic moments under larger magnetic field. When the magnetic field changed from -6 kOe to -10 kOe, the contrast only decreased slightly, in line with the saturation of the magnetization. As the sequence of the applied magnetic field was reversed along red arrow in Fig. 2a, the similar evolution of the magnetic domain was observed. To quantitatively understand the magnetic-field dependence of the MFM images, the magnetic domain contrast (defined as the standard deviation of the image pixels) at the center part of the MFM images has been calculated and plotted against the magnetic field as shown in Fig. 2c. The domain contrast is small at 10 kOe and -10 kOe, which corresponds to the saturation of magnetization associated with a homogeneous single magnetic domain. The maximum contrast occurs at 4 kOe, corresponding to the appearance of magnetic multi-domain states near the coercivity, which is in good agreement with the coercive field of h-YbFeO$_3$ films measured previously [12].

The length scale of the magnetic domains of the h-YbFeO$_3$ films can be identified during the reversal of the magnetization in the MFM images. As shown in Fig. 2d, the average sizes of the magnetic domains increase from 500 nm at -2 kOe to 700 nm at -6 kOe. Therefore, within the $3\ \mu m \times 3\ \mu m$ scan area, the average size of magnetic domains in h-YbFeO$_3$ films is two orders of magnitude smaller than that in h-(Lu,Sc)FeO$_3$ (~ 100 μm) [16]. This is consistent with the enhanced magnetization of h-YbFeO$_3$ which is ~0.1 μ$_B$/f.u. at 12 K [12], about 10 times larger than that of h-(Lu,Sc)FeO$_3$, and the correspondingly enhanced magneto-static energy. The size mismatch between the magnetic domains (~ 100 μm) and the FE domains (~ 1 μm) in h-(Lu,Sc)FeO$_3$ single crystals was taken as an indication of the mismatch between the magneto-static and electrostatic energies, resulting in the bulk ME decoupling [16]. Here in h-YbFeO$_3$ films, due to the enhanced magneto-static energy and the reduced length scale of the magnetic domains, the ME coupling cannot be readily excluded.

**XMCD measurements with in-situ FE polarization switching**

To investigate potential ME coupling in h-YbFeO$_3$ thin films, we first demonstrate the FE and magnetic switching of the h-YbFeO$_3$/CFO/LSMO/STO films. The representative FE polarization versus voltage (*P-V*) loop was measured with 50 Hz frequency at 20 K, as shown in Fig. 3a. The remnant polarization is about 4 μC/cm$^2$, and the coercive voltage is about 4 V [21]. Moreover, the current-voltage relations were also performed using DC voltage at various temperatures, demonstrating clear current peaks corresponding to the FE polarization switching (Fig. S3 of the Supporting Information), which is in good agreement with the coercivity of $P$–$V$ loop in Fig. 3a. To elucidate the magnetic switching, we carried out the XMCD based on x-ray absorption spectroscopy (XAS) under different magnetic fields. The absorption edges of Yb $M_5$ and Fe $L_3$ correspond to the $3d_{5/2}$ to $4f_{7/2}$ in Yb and the $2p_{3/2}$ to $3d$ in Fe, respectively, which correlates to the magnetization of the Yb and Fe sites. The XMCD contrast was defined as $(XAS^+-XAS^-)/(XAS^++XAS^-)$, where $XAS^+$ and $XAS^-$ are the x-ray absorption under or after the application of +18 kOe and -18 kOe magnetic fields. Fig. 3b,c show the representative XAS results at the Yb $M_5$ and Fe $L_3$ edges, respectively. The clear XMCD contrasts can be recognized between the spectra measured under +/- 18 kOe magnetic fields, indicating that the magnetization in the h-YbFeO$_3$ films can be switched using external magnetic fields [12]. Furthermore, clear XMCD contrast from remnant magnetization was observed after +/- 18 kOe magnetic fields were removed, as shown in Fig. 3d, suggesting that the nonvolatile magnetization states can be set by the external magnetic field.

With the FE and ferromagnetic switching demonstrated in h-YbFeO$_3$ films, the ME coupling can be examined. Two approaches have been adopted: A) to set the magnetization state initially, then measure the XAS contrasts between different FE polarization states, B) to change the FE polarization states at first, followed by the measurement of XMCD contrasts using +/- 18 kOe magnetic field.

The approach A) can be described as the following: 1) the initial magnetization state of the h-YbFeO$_3$ films was set by applying and removing an 18 kOe or a -18 kOe magnetic field; 2) the FE polarization state was set by applying and removing a vertical positive DC voltage as shown in Fig. S4 (see Supporting Information), followed by the measurement of x-ray absorption spectra ($XAS^+$);

3) the FE polarization was reversed by applying a negative DC voltage, followed by the measurement of the x-ray absorption spectra (XAS$^-$). The x-ray electric circular dichroism (XECD) was defined as the XAS contrast (XAS$^+$-XAS$^-$) with different FE polarization. The XECD contrasts were measured for both positive (M+) and negative (M-) magnetization states at the Yb $M_5$ edge. As shown in Fig. 3d, the magnitude of XECD contrasts is at least one order of magnitude smaller than that of the XMCD contrast, demonstrating that there is no obvious change of magnetization with the FE polarization switching.

In approach B), the FE polarization states were set by applying and removing in-situ electric fields, followed by the XMCD measurement using the contrast of XAS under +/- 18 kOe magnetic field; this process was repeated with a sequence of applied voltages from 9 V to -9 V to 9 V. For every applied voltage, the XMCD contrasts were measured more than 10 times to reduce the experimental error.

As shown in Fig. 4a, the XMCD contrast for Yb $M_5$ edge exhibits a peak around 1520 eV, there is no obvious shift of peak position associated with various applied voltages (see Fig. S5 of the Supporting Information). The indices for individual XMCD measurements correspond to the number on the in-situ *I-V* curve in Fig. 4b. The clear FE switching current peaks were identified in the in-situ *I-V* curve in Fig. 4b, which is in good agreement with the ex-situ *I-V* curve measured by DC current in Fig. S3 (see Supporting Information) and the *P-V* loop in Fig. 3a. As shown in Fig. 4c, at the Yb $M_5$ edge, the XMCD contrast between negative (+9 V) and positive polarization (-9 V) only exhibits moderate changes. The steepest change of XMCD contrasts occurs near the coercive voltage. Similarly, the voltage-dependent XMCD contrast at the Fe $L_3$ edge shows variations near the coercive voltage. The above results suggest the possible change of magnetization at the coercive voltage, which could be attributed to the domain-wall ME coupling [15, 22]. More thorough in-situ measurements are necessary to fully reveal this effect.

**Theoretical analysis based on Landau theory**

To understand bulk ME decoupling as well as potential domain-wall ME coupling in h-YbFeO$_3$ thin films, a phenomenological model associated with the structural free energy and magnetic energy was considered. As shown in Fig. 5a, the Fe atoms are surrounded by 5 oxygen atoms, forming the trigonal-bipyramid local environments. The structural distortion and the spin configuration can be described using $\phi_Q$ and $\phi_L$, which correspond to the displacement angle of the apex oxygen and the Fe spin orientation in the same FeO$_5$ bipyramid, respectively. The free energy from the Landau theory and magnetic energy can be written as

$$f_1 = \frac{a}{2}Q^2 + \frac{b}{4}Q^4 - gQ^3 P\cos3\phi_Q + \frac{g'}{2}Q^2 P^2 + \frac{a_p}{2}P^2 + \frac{s}{2}\left[(\nabla Q)^2 + Q^2(\nabla \phi_Q)^2\right]$$

$$f_2 = 2S(\partial_u \phi_L)^2 + (A + C_+)\cos2(\phi_L - \phi_Q)$$

Here $f_1$ represents the part of K$_3$ structural distortion (magnitude $Q$), which is coupled with the electric polarization ($P$) [9]. The energy contribution of K$_3$ structural distortion comes from the first two terms, while the third and the fourth terms represent the non-linear coupling between K$_3$ structural distortion and the polar $\Gamma_2^-$ structural distortion whose amplitude is proportional to the electric polarization; the last stiffness terms account for the energy of spatially varied K$_3$ structural distortion. Under the condition of fixed $Q$, the angle $\phi_Q$ corresponds to the collective tilt direction of the FeO$_5$ bipyramids. The equilibrium state, corresponding to the energy minima of $f_1$, locates at $\phi_Q = n\frac{\pi}{3}$. The nonlinear coupling term $gQ^3 P\cos3\phi_Q$ determines the relation between the polarization $P$ and the angle $\phi_Q$, as illustrated in Fig. 5b.

Besides the direct relationship between FE polarization and $\phi_Q$, the magnetization direction along $c$ axis can be determined by $\phi_Q - \phi_L$. Assuming a fixed alignment between the angles $\phi_Q$ (and $\phi_L$) on two FeO$_5$ layers at $z=0$ and $z=c/2$, the magnetic part of the free energy $f_2$ includes two terms. The first term comes from the nearest-neighbor in-plane exchange stiffness; the second term combines the single-ion anisotropy and the interlayer-interaction; the term of Zeeman energy is omitted assuming the absence of an external magnetic field. As shown in Fig. 5c, under the condition $\phi_Q = \pi$, corresponding to a negative polarization, the parallel and antiparallel alignment between $\phi_Q$ and $\phi_L$ would lead to the negative and positive magnetization along the $c$ axis, respectively. Summarizing the above discussion, the FE polarization ($P$) and the magnetization ($M$) could be written as:

$$P = \frac{gQ^3 \cos 3\phi_Q}{g'Q^2 + a_P}, \quad M = -\cos(\phi_Q - \phi_L) M_s,$$

where $M_s$ is the saturation magnetization. For h-YbFeO$_3$ thin films at low temperature, $M_s$ mostly comes from the magnetic moments of Yb ions [12].

Next, we discuss bulk ME coupling/decoupling behavior during the polarization switching using $\phi_Q$ and $\phi_L$ as independent order parameters. First, we note that although the FE polarization reversal can be achieved by changing the K$_3$ structural distortion angle with respect to $\Delta\phi_Q = \frac{\pi}{3}$ or $\Delta\phi_Q = \pi$ [see Fig. 5b], taking account into the high energy barrier, only $\Delta\phi_Q = \frac{\pi}{3}$ is likely to occur [9]. Combining $\phi_Q$ and $\phi_L$ together, the corresponding bulk-state ME coupling or decoupling are illustrated through the landscape of total energy, as shown in Fig. 6a. Given fixed $Q$ with an intermediate value (0.33 Å), the energy minima correspond to the states where $\phi_Q$ equals to $\frac{n\pi}{3}$ ($n=0, 1, 2, 3, 4$ or $5$), as illustrated in Fig. 6b. Meanwhile, to minimize free energy, $\phi_Q - \phi_L = 0$ or $\pi$ is required by the magnetic anisotropy, suggesting that $\phi_L$ would follow the change of $\phi_Q$ during the process of FE polarization reversal. In addition, after the FE polarization reversal (change $\phi_Q$ by $\frac{\pi}{3}$), there are two equivalent energy minima for $\phi_L$. According to the energy landscape in Fig. 6a, for an initial state $\{\phi_Q = \phi_L = 0\}$ with positive FE polarization and the FE switching of $\Delta\phi_Q = \frac{\pi}{3}$, there are two neighboring energy minima ($\{\phi_L = \phi_Q = \frac{\pi}{3}\}$ and $\{\phi_Q = \frac{\pi}{3}, \phi_L = 4\frac{\pi}{3}\}$) with negative FE polarization, suggesting two possible switching paths. However, the energy barrier for the path $\{\phi_Q = \phi_L = 0\} \rightarrow \{\phi_L = \phi_Q = \frac{\pi}{3}\}$ is much lower than the other path, favoring the transition between these two states. As the amplitude of the K$_3$ structural distortion ($Q$) increases, both energy barriers increases, however, the $\{\phi_Q = \phi_L = 0\} \rightarrow \{\phi_L = \phi_Q = \frac{\pi}{3}\}$ barrier always remains lower than that of the other path (see Fig. S7 of the Supporting Information). In Fig. 6b, the coupling and decoupling situations between individual single-domain states with reversed FE polarization are depicted, corresponding to the two paths demonstrated in Fig. 6a. This explains the bulk ME decoupling behaviors demonstrated by the XAS (see Fig. 3 and Fig. 4).

On the other hand, it was previously argued that, despite the higher energy barrier, the $\{\phi_Q = \phi_L = 0\}$ to $\{\phi_Q = \frac{\pi}{3}, \phi_L = \frac{4\pi}{3}\}$ transition may lead to a transition from the single magnetic domain to multi domains resulting in a lower magneto-static energy [8]. Since h-YbFeO$_3$ has larger magneto-static energy than that of h-(Lu,Sc)FeO$_3$, the switching paths for single domain states may be disrupted. To quantitatively understand the influence of magneto-static energy on the switching path, we examined the energy barrier between the two final states $\{\phi_L = \phi_Q = \frac{\pi}{3}\}$ and $\{\phi_Q = \frac{\pi}{3}, \phi_L = \frac{4\pi}{3}\}$. Assuming that the magneto-static energy gained by forming the multiple magnetic domains lowers the barrier, we calculated energy barrier as a function of the saturation magnetization $M_s$. As shown in Fig. S8a (see Supporting Information), when $M_s$ =0, the energy barrier is 0.06 meV/f.u., which decreases quadratically with $M_s$ according to magneto-static energy by $\frac{1}{2}\mu_0 M_s^2$. Specifically, as shown in Fig. S8b (see Supporting Information) the energy barrier decreases to 0.058 meV/f.u when $M_s$ is 1.6 μB/f.u., corresponding to the magnetic moment of Yb observed in h-YbFeO$_3$ films, where the reduction is only 3.9% compared with the $M_s=0$ case. Therefore, the decoupling path $\{\phi_Q = \phi_L = 0\} \rightarrow \{\phi_L = \phi_Q = \frac{\pi}{3}\}$ is still energetically favorable even in h-YbFeO$_3$ since the enhanced magneto-static energy is still far from enough to overcome the energy barrier.

The above analysis explains the bulk ME decoupling assuming that the initial and final states have single FE and magnetic domains. On the other hand, the existence of multiple FE domains and the corresponding domain walls should be also taken into account, especially when the number of FE domain walls proliferates near coercive voltage where the net polarization is zero. Between two neighboring FE domains which have different $\phi_Q$, $\phi_L$ must also be different due to the restriction of single-ion magnetic anisotropy. In other words, FE domain walls should also be antiferromagnetic domain walls. Fig. 6c represents a clamped magnetic domain wall between α$_+$ ($\phi_Q = 0$) and β$_-$ ($\phi_Q = \frac{\pi}{3}$) FE domains. Fig. 6d shows a HAADF image corresponding to a neutral FE domain wall inside h-YbFeO$_3$ film, where the width of FE domain wall ($< 1\ nm$) is far thinner than the magnetic domain wall which is on the order of a few to tens of nanometers. Therefore, near the structural domain wall, $\phi_L$ exhibits slower change than $\phi_Q$, and the magnetization near the FE domain wall is reduced due to the deviation of $\phi_Q$ and $\phi_L$ from their bulk alignment $\phi_Q - \phi_L = 0, \pi,$ based on numerical simulations (see Supporting Information Section S3).

**DISCUSSION**

Previous work showed that the uncompensated magnetic moment at the FE domain wall contributes to the collective magnetism as well as the ME coupling between FE domains and antiferromagnetic domains in h-*R*MnO$_3$ [9, 15, 22]. Here, instead of the generating magnetization at the domain wall in h-*R*MnO$_3$, the magnetization at the FE domain wall is reduced in the h-YbFeO$_3$ films. During the FE switching, when the original FE single-domain state is driven into to the multi-domain state, the increased fraction of the FE domain walls may lead to the observable reduction of average magnetization, corresponding to the ME coupling between FE and magnetic domain walls reflected by the reduced XMCD contrast of Yb M$_5$ and Fe L$_3$ edges near coercive voltage.

In conclusion, the bulk ME decoupling and potential ME coupling of domain wall have been revealed in epitaxial h-YbFeO$_3$ films from experiments and theoretical analysis. The magnetic domain evolution under magnetic field associated with domain distribution (< 1μm) is identified using MFM, and the in-situ measurements of XMCD contrast at Yb M$_5$ and Fe L$_3$ edges illustrate a negligible change of magnetization under single domain states with different FE polarization. Theoretical analysis demonstrates that the ME-decoupling switching path is energy favorable, providing a reasonable explanation for the negligible change of XMCD contrasts under different applied voltage; the potential reduction of XMCD contrast near coercive voltage may originate from the suppressed magnetization at FE domain wall according to the theoretical calculations. Our results offer fundamental insights into ME coupling and decoupling process in improper multiferroic hexagonal ferrites and pave the way for the future applications of multiferroics, such as micro-sensors, micro-electro mechanical systems (MEMS) and high-density information storage [23].

**METHODS**

**Sample preparation**

The h-YbFeO$_3$ thin films (20-100 nm thick) were grown on CoFe$_2$O$_4$ /La$_{2/3}$Sr$_{1/3}$MnO$_3$ / SrTiO$_3$ (111) and yttrium stabilized zirconia (YSZ) (111) substrates by pulsed laser deposition (PLD) system with a KrF excimer laser (248 nm and 2 Hz repetition rate), at the growth temperature from 650 °C to 850 °C and oxygen pressure of 10 mTorr. Before thin-film deposition, substrates were pre-annealed at 700 °C for 1 hour. The $La_{2/3}Sr_{1/3}MnO_3$ (LSMO) layer ($\approx$ 30 nm) was grown at a substrate temperature of 700 °C and oxygen pressure of 80 mTorr on the $SrTiO_3$ (STO) substrate. The $CoFe_2O_4$ (CFO) layer ($\approx$ 10 nm) was grown at the temperature of 600 °C and the oxygen pressure of 10 mTorr. The film growth was monitored using in-situ reflection high-energy electron diffraction (RHEED). The Au (3-5 nm) top electrodes were evaporated by an AJA sputtering system with 300-400 μm diameter.

**Structural characterization**

The structural phase of the epitaxial films was determined using X-ray diffraction (XRD) (Rigaku SmartLab). Scanning transmission electron microscopy (STEM) imaging was carried out using the aberration-corrected Nion UltraSTEM™ 200 microscope (operating at 200 kV) at Oak Ridge National Laboratory. An electron transparent thin foil for STEM characterization was prepared using a Hitachi NB5000 focused ion and electron beam system. To protect against the ion beam damage, a 1-μm-thick carbon layer was deposited on top of the h-$YbFeO_3$ film surface. A 20 kV beam with a current of 0.7 nA was used to cut the lift-out. Rough and fine milling were performed at 10 kV and 5kV with beam currents of 0.07 nA and 0.01 nA respectively. The resulting foil was mounted on a Cu grid, which was baked at 160 °C under vacuum prior to the STEM experiments to remove surface contamination.

**Magnetic domain measurements**

The atomic force microscopy (AFM) and magnetic force microscopy (MFM) images were performed at low temperature using an Attocube system.

**XMCD measurements**

The X-ray absorption spectroscopy (XAS) (including X-ray magnetic circular dichroism or XMCD) was studied at the beamline 6.3.1 in the Advanced Light Source at Lawrence Berkeley National Laboratory.

**Ferroelectric measurements**

The FE polarization was switched by the DC current using a Keithley 236 source meter (the measurements of high-resolution *I-V* curves) and a Keithley 2450 source meter (in-situ polarization switching during XMCD measurements). The polarization versus electric filed (*P-V*) loops were measured using a Precision RT66C Ferroelectric Tester.

**Data availability**

All relevant data are available from the authors upon request.

**Acknowledgements**

**Funding:** This work was primarily supported by the National Science Foundation (NSF), Division of Materials Research (DMR) under Grant No. DMR-1454618. The research was performed in part in the Nebraska Nanoscale Facility: National Nanotechnology Coordinated Infrastructure and the Nebraska Center for Materials and Nanoscience, which are supported by the NSF under Grant No. ECCS- 2025298, and the Nebraska Research Initiative. Work at Washington University (electron microscopy) was supported by NSF Grant No. DMR-1806147. STEM experiments were conducted at the Center for Nanophase Materials Sciences at Oak Ridge National Laboratory, which is a Department of Energy (DOE) Office of Science User Facility, through a user project.

**Author contributions**

X.X. conceived and designed the experiments. The materials are fabricated by Y.Y. with the assistance from X.X. and Y.Yin. The MFM experiments were conducted and analyzed by X.X. with the assistance from H.L., Q.L. and W.W. The XMCD experiments conducted by Y.Y., C.M. and X.J. with the assistance from A.T.N. and X.X. The analysis of XMCD were conducted by Y.Y. and X.L. The theoretical calculations were carried out by X.L. under the guidance of X.X. The XRD experiments and ferroelectric properties were performed by Y.Y. STEM experiments were conducted by A.S.T. under the guidance of R.M. Y.Y., X.L. and X.X. wrote the manuscript. All authors discussed the results and commented on the manuscript.

**Additional Information**

Supplementary information

Section S1. Ferroelectric characterization

Section S2. X-ray magnetic circular dichroism (XMCD) measurements

Section S3. ME coupling between ferroelectric domain wall and magnetic domain wall

Fig. S1. RHEED images along STO [1-10]/ (111) direction

Fig. S2. Additional MFM images

Fig. S3. Ex-situ I-V curves of h-YbFeO3 films

Fig. S4. Schematic diagram of XMCD measurement setup

Fig. S5. The shift of XMCD peak position under different applied voltage

Fig. S6. XMCD contrast of Fe L3 edge with applied external voltage

Fig. S7. Energy landscape with various Q

Fig. S8. Energy landscape and energy barrier under different Ms

**Competing Interests**

The authors declare that they have no competing interests.

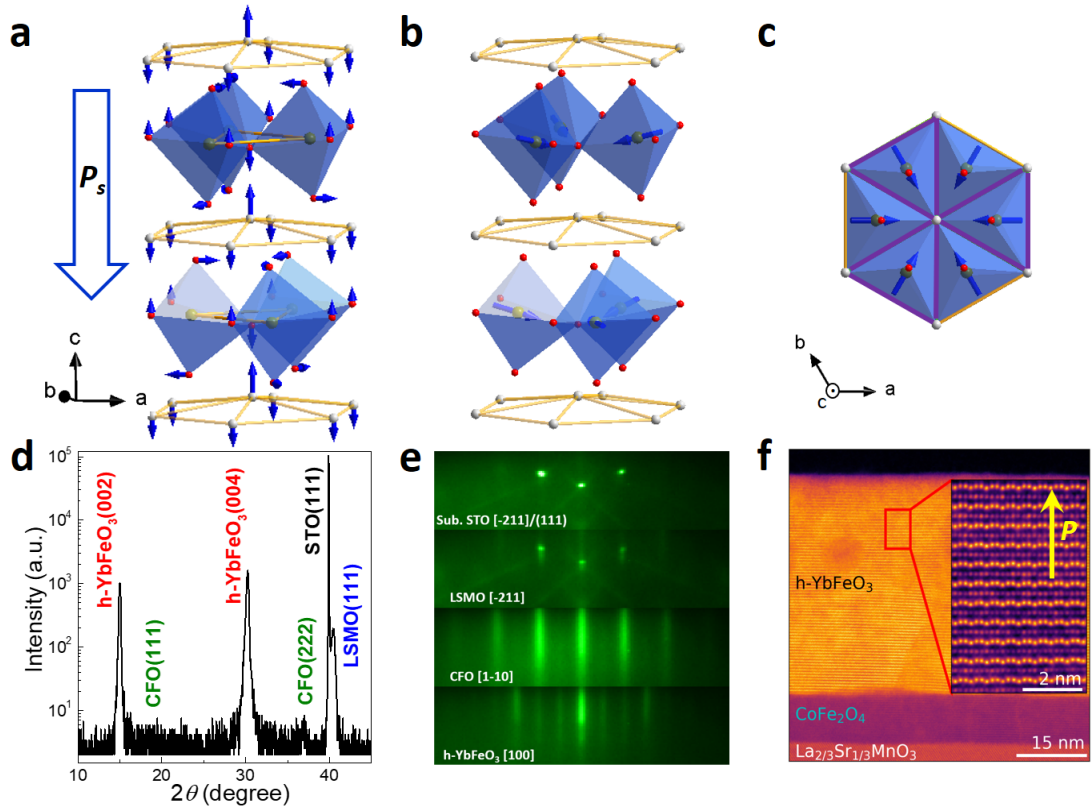

**Fig. 1** Atomic structure of h-$R$FeO$_3$ and structure characterization of the h-YbFeO$_3$/CFO/LSMO/STO heterostructure. (**a**) Atomic structure and K$_3$ structural distortion of h-$R$FeO$_3$. Brown, white, and red spheres represent Fe, rare-earth, and O atoms, respectively. The arrows indicate the displacement of O and Yb atoms of the K$_3$ mode. (**b**) Schematic illustration of WFM of h-$R$FeO$_3$, the arrows through the Fe atoms represent the spin directions. (**c**) The in-plane spin alignment of Fe atoms. The FeO$_5$ bipyramids at $z$=c/2 layer are highlighted by the purple triangles. (**d**) $\theta$-$2\theta$ XRD measurement of h-YbFeO$_3$/CFO/LSMO/STO films. (**e**) RHEED images of h-YbFeO$_3$/ CFO/ LSMO/ STO (111) heterostructure along STO [-211]/(111) direction. f) HAADF image of h-YbFeO$_3$/CFO/LSMO layers viewed along h-YbFeO$_3$ [100] direction. The magnified image corresponds to the red box area, which shows a periodic "two-up-one-down" distribution for Yb atoms (the brightest atoms), indicating FE polarization pointing up.

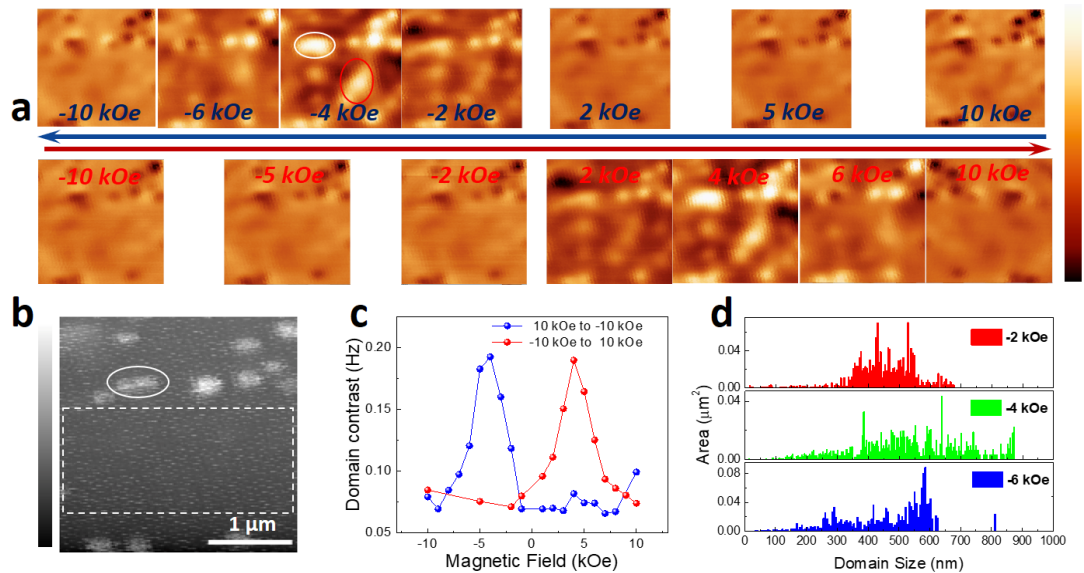

**Fig. 2** Magnetic domains performed by MFM. (**a**) MFM images of h-YbFeO$_3$(001)/ YSZ (111) were measured in the phase-lock mode, 100 nm above the surface, with a 3×3 μm$^2$ scan area. The gray scale of the images corresponds to the frequency shift, whose full scale is 1 Hz. The arrows indicate the sequence of the applied magnetic field along the *c* axis of h-YbFeO$_3$ films. Representative defect sites and nucleation sites are labeled by white and red circles, respectively. (**b**) Surface topography of h-YbFeO$_3$ films measured by AFM with a 75 nm full grey scale. Representative defect site (the island) is labeled by white circle. The dashed rectangle represents the region without islands. (**c**) Magnetic-field dependence of the magnetic domain contrast calculated using the central part of the images in **a**, corresponding to dashed square region in **b**. (**d**) The distribution of magnetic domain size extracted from the central part of the images in **a** with various magnetic field. All measurements were taken at 12 K.

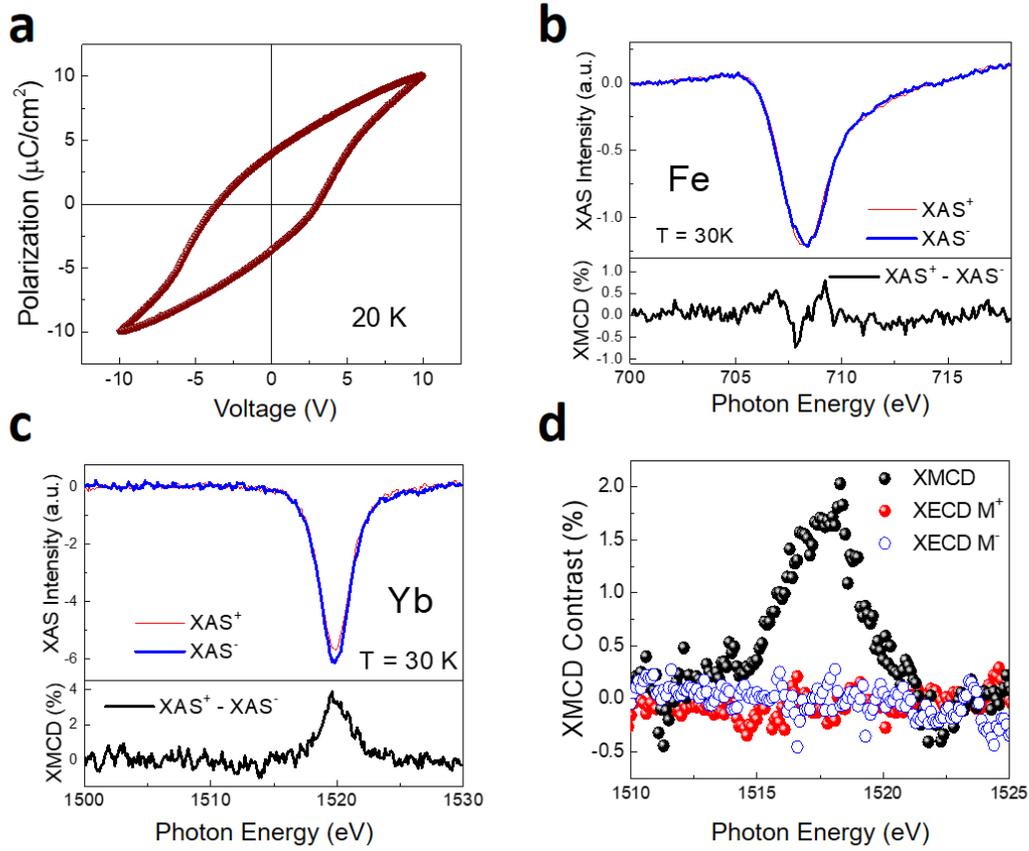

**Fig. 3** Representative hysteresis loop and XMCD contrast for h-YbFeO$_3$ films. (**a**) Representative FE hysteresis loop of the h-YbFeO$_3$ (34nm)/CFO/LSMO/STO film at 20K. (**b**) and (**c**) are the typical XAS$^+$ (under +18 kOe), XAS$^-$ (under -18 kOe), and XMCD contrast (XAS$^+$-XAS$^-$)/(XAS$^+$+XAS$^-$) measured at 30 K, at the Fe L$_3$ edge and the Yb M$_5$ edge, respectively. (**d**) XMCD contrasts of the remnant magnetization (measured under zero magnetic field after application of ± 18 kOe magnetic field) and X-ray electric circular dichroism (XECD) contrasts between FE polarization up and down under magnetic field +18 kOe (XECD M$^+$) and -18 kOe (XECD M$^-$) at Yb M$_5$ edge.

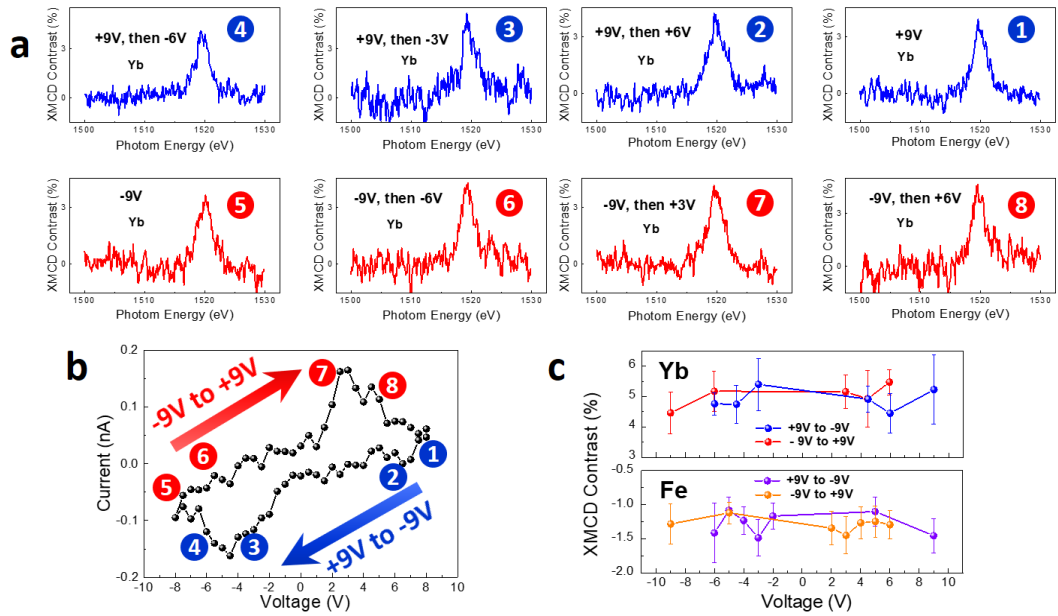

**Fig. 4** XMCD measurements with in-situ ferroelectric switching. (**a**) XMCD contrast of Yb at $M_5$ edge with applied external voltage from 9 V to -9 V (blue) and from -9 V to 9 V (red). (**b**) In-situ measurement of *I-V* curve during the process of measuring XMCD at 30 K. (**c**) Voltage dependence of XMCD contrasts at the Yb $M_5$ edge and Fe $L_3$ edge.

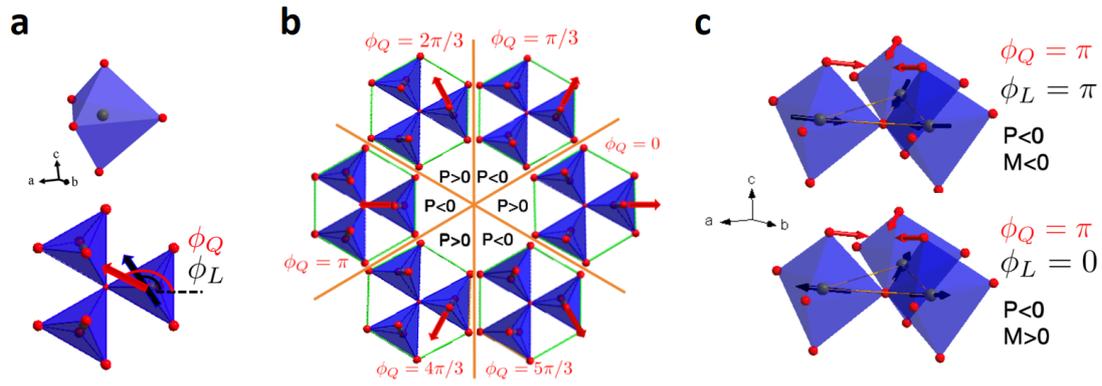

**Fig. 5** Schematic diagram of domains and crystal structure. (**a**) Schematic diagram of FeO$_5$ bipyramids and definition of two order parameters corresponding to the rotation angle of apex oxygen ($\phi_Q$) and spin direction angle of Fe ($\phi_L$), red and grey spheres represent O and Fe atoms. (**b**) Different polarization states with $\phi_Q$ from 0 to $5\pi/3$. (**c**) Schematic diagram of polarization ($P$) and magnetization ($M$) with different directions corresponding to relative angle of apex oxygen ($\phi_Q$) and spin of Fe ($\phi_L$).

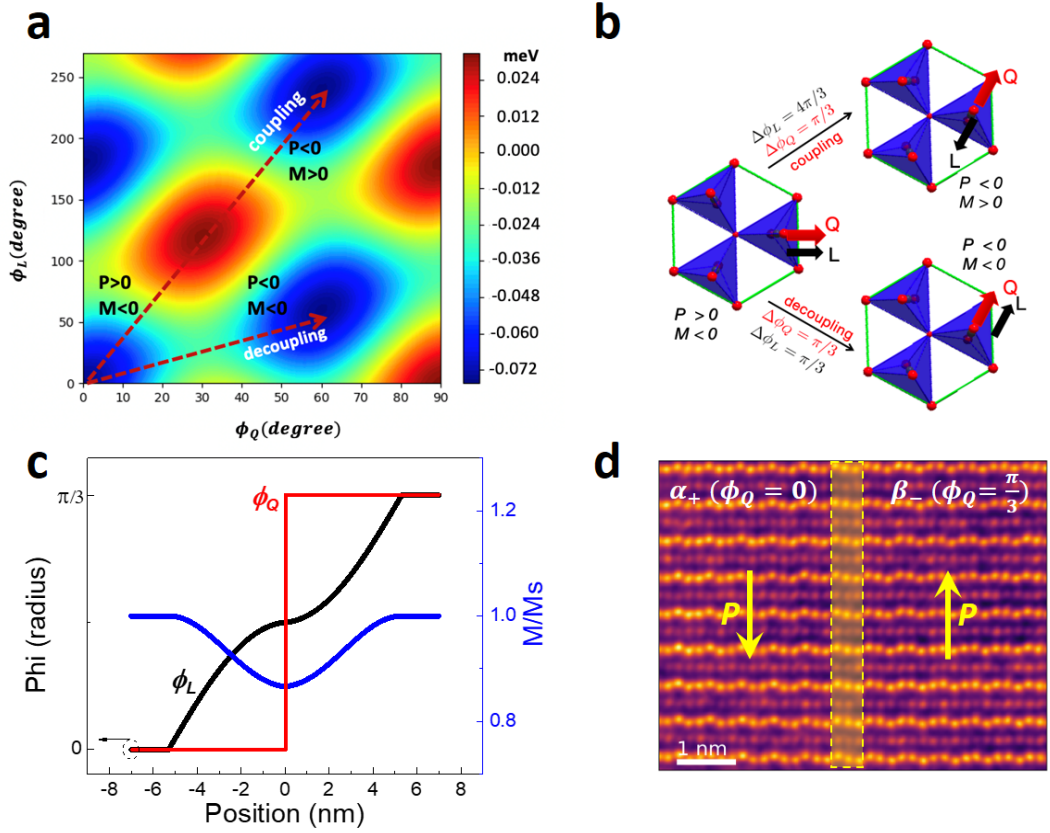

**Fig. 6** Bulk ME decoupling and domain-wall ME coupling revealed by Landau theory. (**a**) Energy landscape with $\phi_Q$ and $\phi_L$ as coordinates, where $Q$ is fixed at 0.33 Å. (**b**) Schematic illustration of ME coupling and decoupling situations during the process of FE polarization switching. (**c**) Magnetic domain wall clamped to the FE domain wall between the $\phi_Q = 0$ domain and the $\phi_Q = \pi/3$ domain. (**d**) HAADF image showing a neutral domain wall between $\alpha_+(\phi_Q = 0)$ and $\beta_-(\phi_Q = \pi/3)$ FE domains in h-YbFeO$_3$ thin film, viewing from <100> direction. The domain wall width is less than 1 nm.

**Supplementary Material**

**Magnetoelectric coupling and decoupling in multiferroic hexagonal YbFeO$_3$ thin films**

*Yu Yun,[1][‡] Xin Li,[1][‡] Arashdeep Singh Thind,[2] Yuewei Yin,[1] Hao Liu,[3] Qiang Li,[3] Wenbin Wang,[3] Alpha T. N'Diaye,[4] Corbyn Mellinger,[1] Xuanyuan Jiang,[1] Rohan Mishra,[2,5] Xiaoshan Xu[1,6]\**

[1]Department of Physics and Astronomy, University of Nebraska, Lincoln, Nebraska 68588, USA

[2]Institute of Materials Science & Engineering, Washington University in St. Louis, St. Louis MO, USA

[3]Institute for Nanoelectronic Devices and Quantum Computing, Fudan University, Shanghai 200433, China

[4]Advanced Light Source, Lawrence Berkeley National Laboratory, Berkeley, California 94720, USA

[5]Department of Mechanical Engineering & Materials Science, Washington University in St. Louis, St. Louis MO, USA

[6]Nebraska Center for Materials and Nanoscience, University of Nebraska, Lincoln, Nebraska 68588, USA

**Section S1. Ferroelectric characterization**

Fig. S3 shows *I-V* curves at various temperature measured by Keithley 2450 source meter using DC current. The voltage step is 0.4V. The *I-V* curves demonstrate obvious current peaks due to ferroelectric polarization switching, which is consistent with the coercive voltage of the *P-V* loop (Fig. 3a) and the in-situ *I-V* curves (Fig. 4b), suggesting the ferroelectric polarization can be controlled by the DC current during the measurements of XMCD.

**Section S2. X-ray magnetic circular dichroism (XMCD) measurements**

Fig. S4 shows the schematic diagram of XMCD measurement setup. The top electrodes and bottom electrodes are connected to the source meter using silver paint. The diameter of top electrodes is from 300 μm to 400 μm, which can provide enough area for the measurements of XMCD. The silver paint and top electrodes can reduce the signal of XMCD, especially the thickness of silver paint is from several micrometers to tens of micrometers. Hence, we attached silver paint only on the corner of top electrodes, and the thickness of Au top electrodes only 3-5 nm. Then the beam spot was focused on the top electrode without the silver paint. During the XMCD measurements, the ferroelectric polarization can be in-situ switched by external voltage.

Under specific applied voltage, the XMCD contrasts were measured 8 times in single measurement, and every voltage corresponds to related ferroelectric state on ferroelectric hysteresis loop. Three XMCD contrast and the shift of peak position for Yb M5 edge and Fe L3 edges were shown in Fig. 4c and Fig. S5, respectively.

**Section S3. ME coupling between ferroelectric domain wall and magnetic domain wall**

Considering the fixed angle between in-plane Fe spins at odd and even layers with ($\phi_{L1} - \phi_{L2} = \pi/3$), the magnetic free energy of one-unit cell can be expressed as:

$$f_{mag}(\phi_L, \phi_Q) = 2S(\partial_u \phi_L)^2 + (A + C_+)\cos\left(2(\phi_L - \phi_Q)\right) + A - C_- \quad (1)$$

in which the $\cos\left(2(\phi_L - \phi_Q)\right)$ term combines the magnetic free energy with structural free energy together. Using the $\phi_L - \phi_Q$ as independent variable and applying Euler-Lagrange equation:

$$\nabla \cdot \frac{\partial f_{mag}}{\partial(\nabla(\phi_L-\phi_Q))} = \frac{\partial f_{mag}}{\partial(\phi_L-\phi_Q)} \quad (2)$$

The one-dimensional magnetic domain wall (along x direction) corresponds to:

$$4S\frac{d^2}{dx^2}(\phi_L - \phi_Q) = -2(A + C_+)sin\left(2(\phi_L - \phi_Q)\right) \quad (3)$$

Therefore $(\phi_L - \phi_Q)$ follows the non-linear pendulum equation (3). The analytical result for $\phi_Q(x)$ has been derived in previous research using similar method [18]:

$$\phi_Q(x) = \phi_0 + \frac{2}{3}arctan(e^{x/\xi_6}) \quad (4)$$

in which $\xi_6$ is the characteristic length for FE domain wall, determined by parameters in structural free energy. Since the width of FE domain wall is much smaller than the magnetic domain wall, the spatial variation of $\phi_Q$ can be approximated as step function:

$$\phi_Q = 0, x < 0 \quad and \quad \phi_Q = \frac{\pi}{3}, x > 0 \quad (5)$$

corresponding to neutral domain walls between $\alpha_+$ and $\beta_-$ FE domains. Therefore, the equation (3) can be simplified to:

$$\frac{d^2}{dx^2}(2\phi_L) = -\frac{A+C_+}{2S} sin(2\phi_L), \quad x < 0$$

$$\frac{d^2}{dx^2}(2\phi_L - \frac{2\pi}{3}) = -\frac{A+C_+}{2S} sin(2\phi_L - \frac{4\pi}{3}), \quad x > 0 \quad (6)$$

Basing on equation (6) with continuous boundary condition at x=0 ( $\phi_L|_{0-} = \phi_L|_{0+}$), the spatial variation of $\phi_L$ across sharp FE domain wall could be numerical solved through modified Euler methods. Combing with

$$M = -\cos(\phi_Q - \phi_L)M_s \qquad (7)$$

the profile of weak ferromagnetism around FE domain wall can be simulated, as shown in Fig. 6c.

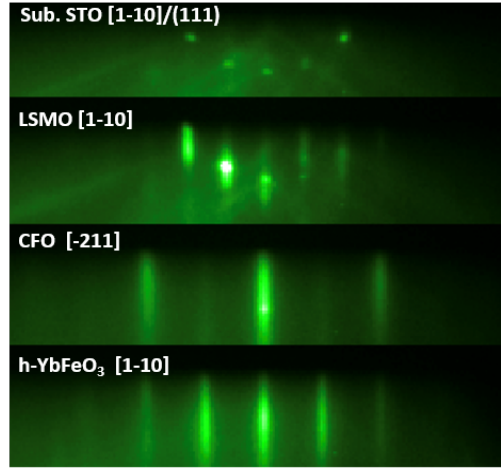

**Fig. S1** RHEED images along STO [1-10]/ (111) direction. RHEED images of the h-YbFeO$_3$/ CFO/ LSMO/ STO (111) heterostructure along STO [1-10]/ (111) direction.

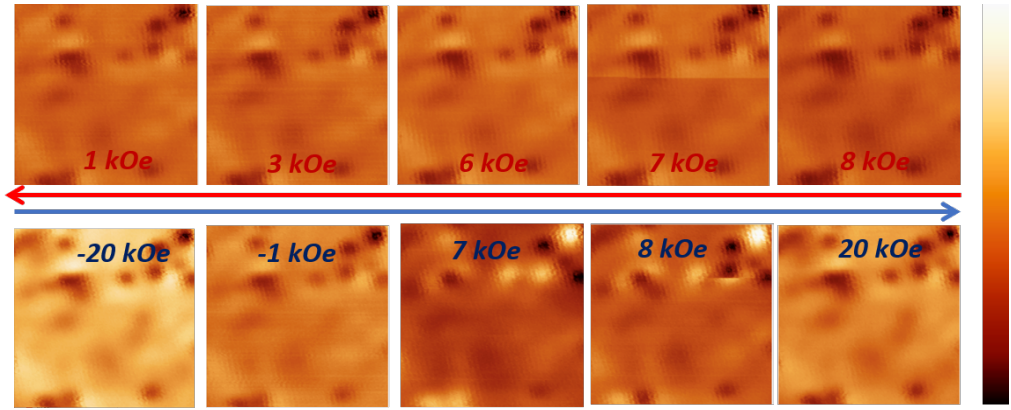

**Fig. S2** Additional MFM images. MFM images of h-YbFeO$_3$ (001)/YSZ under different magnetic field at 12K, corresponding to Fig. 2a. The gray scale of the images corresponds to the frequency shift, whose full scale is 1 Hz.

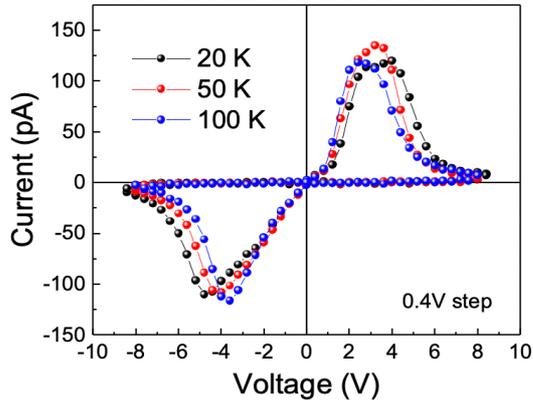

**Fig. S3** Ex-situ *I-V* curves of h-YbFeO$_3$ films. The measurements of *I-V* curve of h-YbFeO$_3$ film using DC voltage at 20 K, 50 K and 100 K.

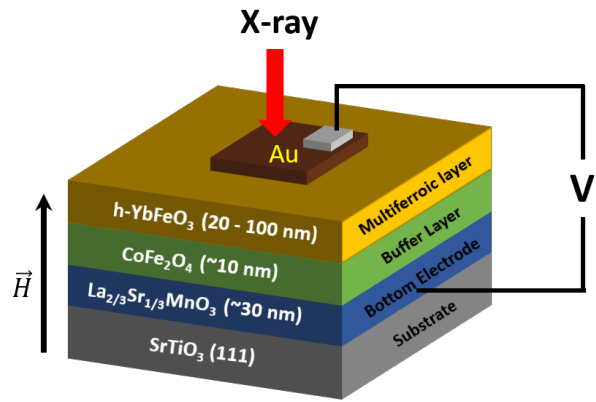

**Fig. S4** Schematic diagram of XMCD measurement setup.

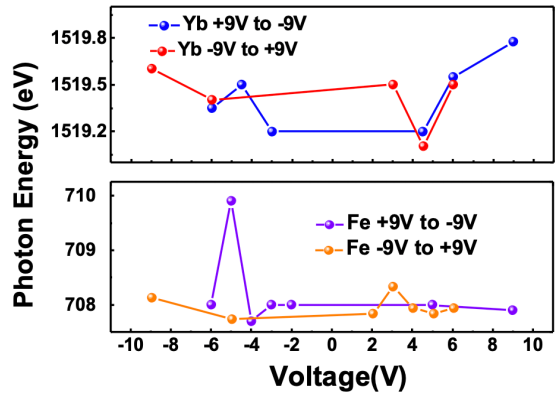

**Fig. S5** The shift of XMCD peak position under different applied voltage. The shift of peak positions for Yb $M_5$ and Fe $L_3$ edges under different applied voltage.

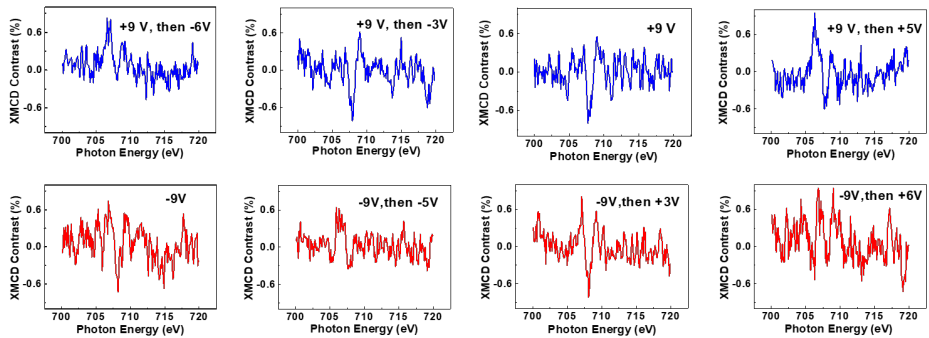

**Fig. S6** XMCD contrast of Fe $L_3$ edge with applied external voltage. XMCD contrast of Fe $L_3$ edge with applied external voltage from 9 V to -9 V (blue) and from -9 V to 9 V (red).

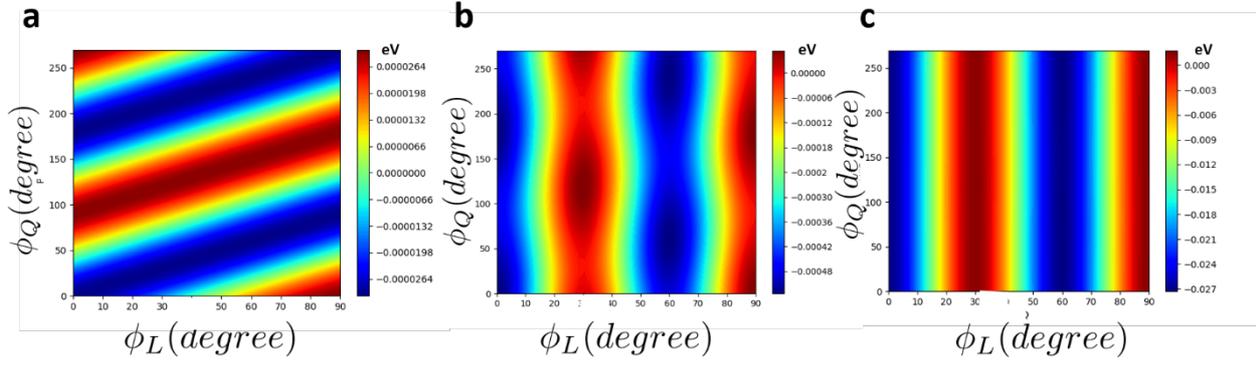

**Fig. S7** Energy landscape with various Q. Energy landscape with Q is fixed at (**a**) 0.1 Å, (**b**) 0.5 Å, and (**c**) 1.0 Å, respectively.

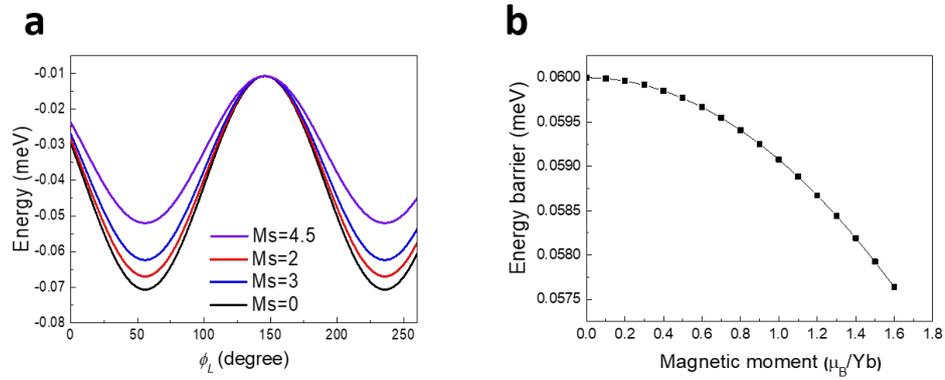

**Fig. S8** Energy landscape and energy barrier under different Ms. (**a**) The change of energy landscape at $\phi_Q = \pi/6$ with saturated magnetization (Ms), the unit of Ms is $\mu_B/Yb$. (**b**) The change of energy barrier between $\phi_L = \pi/3$ and $\phi_L = 4\pi/3$ under different Ms.